\documentclass[aps,prl,twocolumn,amsfonts]{revtex4}
\usepackage{graphicx}

\newcommand{\eq}[1]{Eq.~(\ref{#1})}
\def\etall{{\em et al.\ }}

\begin{document}


\title{Vertex Intrinsic Fitness: How to Produce Arbitrary Scale-Free
Networks}


\author{Vito D.P. Servedio}
\affiliation{INFM UdR Roma1 and Dipartimento di Fisica, Universit\`a ``La
	     Sapienza'', Piazzale A.\ Moro 2, I-00185 Roma, Italy}

\author{Paolo Butt\`a}
\affiliation{ Dipartimento di Matematica, Universit\`a di Roma ``La Sapienza'',
	      Piazzale A.\ Moro 2, I-00185 Roma, Italy}

\author{Guido Caldarelli}
\affiliation{INFM UdR Roma1 and Dipartimento di Fisica, Universit\`a ``La
	     Sapienza'', Piazzale A.\ Moro 2, I-00185 Roma, Italy}

\date{September 29, 2003}

\begin{abstract}
We study a recent model of random networks based on the
presence of an intrinsic character of the vertices called fitness.
The vertices fitnesses are drawn from a given probability
distribution density. 
The edges between pair of vertices  are drawn according to a 
linking probability function depending on the fitnesses of the two 
vertices involved.
We study here different choices for the  probability distribution densities and
the linking functions.
We find that, irrespective of the particular choices, the generation of
scale-free networks is straightforward. 
We then derive the general conditions under which scale-free behavior
appears. 
This model could then represent a possible explanation for the
ubiquity and robustness of such structures.
\end{abstract}


\maketitle



In the last few years, much attention has focused on the study of
complex networks.
A network is a mathematical object consisting of a
collection of vertices (nodes) connected by edges
(links)\cite{Bollobas01,DM03book}.
Networks arise in many areas of science: biology
\cite{KKKR02,bio,bio2}, social
sciences \cite{ASBS00,social1,social2},
Internet \cite{int1,int2,int3}, WWW\cite{AH00}, etc., where vertices and
links can be for example, proteins and their mutual interaction,
individuals and sexual relationship \cite{Liljeros01},  computers and
cable connections.
Very interestingly the same non trivial statistical properties appear
ubiquitously in all the above situations.
A more traditional view, indeed, is represented by the binomial model
inspired to the random graph model of 
Erd\H{o}s-R\'enyi \cite{ER61}.
Here, each vertex has the same
probability to connect to any other, resulting in a network whose
degree, i.e.\ number of edges per vertex, has a binomial distribution.
This is not the case of the above real data, where instead, the
structure is self similar resulting in a scale-free (SF) probability
distribution for the degree.
More specifically, 
the degree $k$ of the vertices, i.e.\ the number of links entering them,
is distributed according to a power law $P(k)\propto k^{\alpha}$ with
usually $-3<\alpha<-2$.

In order to explain the occurrence of SF networks the ingredients of growth
preferential attachment have been introduced \cite{bara}.
The network increases the number of vertices with time, the newcomers
tend  to be connected with old vertices with large degree.
Nevertheless, in some cases, we have the same SF properties without either
growth of the system or preferential attachment mechanism.
As an example, the finite set of protein interactions in a cell forms
a self-similar network.
This is done without growth of the system size and ignoring their
reciprocal degree. 
Possibly, some external influence on intrinsic properties like
chemical affinity is instead driving the phenomenon.

To take into account this new mechanism, the varying fitness
model has been introduced by Caldarelli \etall\cite{gcalda}.
In this model, considering e.g.\ only undirected graphs, one extracts a real
non-negative variable $x$ (the hidden variable) for each vertex of the
graph from a probability distribution density $\rho(x)$.
This variable $x$ is the {\em fitness} of the vertex.
Links between vertices are successively formed with a probability
function $f(x,y)$, a symmetric function of its arguments.

A static simplified form of the vertex hidden variable model has been 
considered for only one particular case by Goh \etall\cite{goh01}.
Bianconi \etall\cite{BB01b} introduced a fitness mechanism coupled to
the preferential attachment.
In the paper of Caldarelli \etall\cite{gcalda},
the onset of SF behavior is instead directly related only
to the fitness presence of any kind.
This behavior is also checked for different fitness probability
distribution densities.
Following Ref.~\cite{gcalda}, we present here an exhaustive study on
the conditions needed in order to produce a SF network.

The aim of this work is to provide some ingredients to generate
SF networks with the vertex hidden variable model and to provide the
analytic expressions for the functions $\rho(x)$ and $f(x,y)$ that
define SF networks in three special cases.


The fitness model can be easily generalized in order to have more than
one fitness variable per vertex \cite{PScondmat}. 
In the following, we consider a single real variable
$x$ per vertex, with $x\ge0$.
As a probability distribution density function, $\rho$ satisfies
$\{\rho(x)\ge0|\int_0^\infty \rho(z) dz = 1\}$, while the linking
probability $0\le f(x,y) \le 1$.
We define the primitive function of $\rho(x)$, the probability
distribution $R(x)=\int_0^x \rho(z)
dz$.
Indicating the number of vertices in the graph with $N$, one has the vertex
degree 
\begin{equation}
	k(x)=N \int_0^\infty \!\!\! f(x,z) \rho(z) dz. \label{k}
\end{equation}
Other quantities of interest are the average nearest neighbor connectivity
(vertex degree correlation),
\begin{equation}
	K_{\mathrm{nn}}(x) = \frac{N}{k(x)} \int_0^\infty \!\!\!
		f(x,z) k(z) \rho(z) dz,
	\label{Knn} 
\end{equation}
expressing the average degree of vertices that are nearest neighbors of
vertices with fitness $x$,
and the clustering coefficient (vertex transitivity),
\begin{equation}
	C(x) = N^2\frac{\int_0^\infty \!\!\! \int_0^\infty \!\!\!
		f(x,y) f(y,z) f(z,x) \rho(y)\rho(z) dy dz}
		{k(x)^2},
	\label{C} 
\end{equation}
that counts the fraction of nearest neighbors of vertices with fitness
$x$ that are also nearest neighbors each other.
Eqs.~(\ref{k}), (\ref{Knn}), (\ref{C})  are  valid asymptotically when $N$
approaches infinity.
Eqs.~(\ref{Knn}), (\ref{C}) were first derived in Ref.~\cite{BP03},
and expressed in a different form.\\
If $k(x)$ is an invertible and increasing function of $x$ then the
probability distribution $P(k)$ is given by
\begin{equation}
	P(k)=\rho(x(k)) \cdot x'(k) \label{Pk}
\end{equation}
or, as a function of $x$,
\begin{equation}
	P(k(x))=\frac{\rho(x)}{k'(x)} \label{Pkx}.
\end{equation}
Since the degree probability is power-law distributed in most of the
physical situations, we impose in \eq{Pkx}
$P(k)=ck^\alpha$ with $\alpha\in\mathbb{R}$. The constant $c$ is fixed
by the the normalization condition \(\int_{k_0}^N P(k) dk =1\):
\begin{equation}
    c = \left\{
	\begin{array}{ll}
		\frac{\alpha+1}{N^{\alpha+1}-k_0^{\alpha+1}} &\text{if } \alpha\neq-1\\
		\left(\log \frac{N}{k_0}\right)^{-1} & \text{if } \alpha=-1\\
	\end{array} \right. \label{cnorm}
\end{equation}
with \(\displaystyle k_{0}=\lim_{x\rightarrow+0} k(x)\). 
Note that $k_0 = \beta N$ for some $0<\beta<1$, and $c \propto
N^{-(\alpha+1)}$.
\eq{Pkx} becomes:
\begin{equation}
	c k'(x) (k(x))^\alpha = \rho(x). \label{eq0}
\end{equation}
By integrating \eq{eq0} from $0$ to $x$ we get the following non linear
integral equation:
\begin{equation}
k(x) = \left\{
	\begin{array}{ll}
		\Bigl(k_0^{\alpha+1} + \frac{\alpha+1}{c} R(x)\Bigr)^
		{\frac{1}{\alpha+1}} & \mathrm{if}\ \alpha \neq -1\\
		k_0 e^{ R(x)/c} & \mathrm{if}\ \alpha=-1\\
		\end{array} \right.
	\label{Ieq}
\end{equation}
with $k(x)$ given by \eq{k}.

By multiplying both sides of \eq{Ieq} by $\rho(x)$ and integrating from $0$ to
$\infty$ we get an analytic expression for the average vertex degree
$\langle k\rangle$. This expression can be used to express $k_0$ as a
function of $\langle k\rangle$, so that the final expressions do
depend on the physical quantity $\langle k\rangle$ only.
For this purpose, the integral on the rhs is simply solved
using the relation $\rho(x) dx=dR(x)$.

In the following we show an application of the model in three
special cases of interest, comparing the analytic results with
numerical simulations.
It has to be noticed that once fixed $N$, in order to compute
the quantities $P(k)$, $K_{\mathrm{nn}}(k)$, $C(k)$, 
from ensemble 
statistics, we need to perform two different average procedures.
Firstly, we should extract a $\{x_i\}_{i=1\ldots N}$ configuration with  the
distribution density $\rho(x)$ and keep it fixed, while creating
ensemble elements using the linking probability $f(x,y)$ and averaging
at the end. 
Secondly, we should repeat the above procedure a sufficient number of
times. 
We assume that for large enough $N$ and ensemble elements, the
procedure of first averaging with respect to the $f$ can be skipped.\\
Here we focus on two different problems:
firstly, what we call direct problem, one assigns a distribution density
function
$\rho(x)$ and tries to find the linking probability function $f(x,y)$;
secondly, what we call inverse problem, one assigns the linking
probability function and tries to determine the fitness probability
distribution density $\rho(x)$.
The inverse problem is by far more complex and interesting than the
direct one. 
For instance, in the case of protein SF network by assuming a
reasonable linking function, we can retrieve the probability density
distribution of fitness (e.g.\ some basic property of the
macro-molecules).
%

\begin{figure}
  \includegraphics[width=7cm]{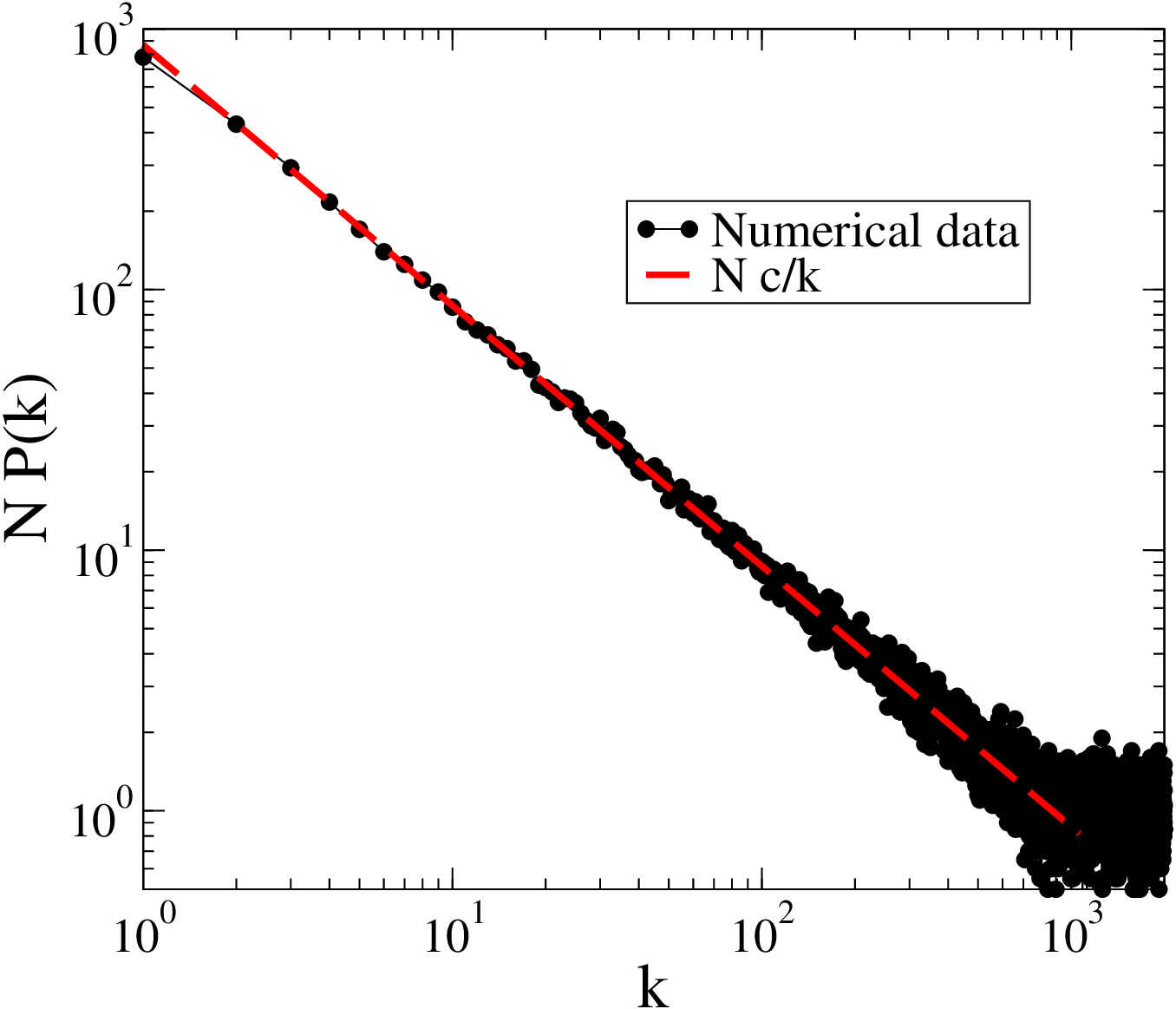}%
  \caption{Vertex degree distribution generated by $f(x,y)=g(x)g(y)$,
	   $\alpha=-1$, $\rho(x)=e^{-x}$, $k_0=0.1$, $N=10^4$.
	   The value of $c$ is calculated from \eq{cnorm} and the
	   function $g(x)$ from \eq{a-1}.
	   The number of ensemble elements was $20$.
	   \label{fig:ck-1}}
\end{figure}

We start with the special case of $f(x,y)=g(x)h(y)$ where both the direct
and inverse problems can be  analytically solved.
Because of the symmetry of $f(x,y)$ with respect to its arguments one has
$g(x)\equiv h(x)$, so that $f(x,y)=g(x)g(y)$.
\eq{k} becomes:
\begin{equation}
	k(x) = N g(x) \int_0^\infty \!\!\! g(z) \rho(z) dz \label{kgg}
\end{equation}
that substituted into \eq{Ieq} gives equations in $g$ and $\rho$.
If one fixes a given function $\rho(x)$, the equations in $g(x)$
can be easily solved.
Take for instance the second equation corresponding to $\alpha=-1$.
One gets:
\[
	N g(x)\langle g \rangle = k_0 e^{R(x)/c} , \quad \langle g \rangle=\int_0^\infty \!\!\!
		g(z)\rho(z) dz .
\]
By multiplying left and right hand side by \(\rho(x)\) and integrating
from $0$ to $\infty$, considering that \(\rho(x)dx = dR(x) \), we get:
\[
	\langle g \rangle = \sqrt{k_0 c (e^{1/c}-1)/N}.
\]
Finally, the solution reads:
\begin{equation}
	g(x) = \sqrt{\frac{k_0}{N c (e^{1/c}-1)}} \, e^{R(x)/c}.\label{a-1}
\end{equation}
This procedure is applicable for any value of $\alpha$.
\eq{a-1} generates random networks with degree
probability distribution $P(k)\propto 1/k$.\\
In order to test the result we take the choice reported in the caption
of Fig.~\ref{fig:ck-1}.
We conclude that {\em for any given $\rho(x)$ there exists a function $g(x)$ such
that the network generated by $\rho(x)$ and $f(x,y)=g(x)g(y)$ is scale
free with arbitrary real exponent}.

In this case both the average nearest neighbors connectivity
and clustering coefficient are constant \cite{BP03}.
Respectively:
\begin{equation}
	K_\mathrm{nn} = N\langle g^2 \rangle, \quad
	C = \frac{\langle g^2 \rangle^2}{\langle g \rangle^2},
\end{equation}
as it can be derived from \eq{Knn} and \eq{C}.

The inverse problem for $f(x,y)=g(x)g(y)$ is solved
by substituting \eq{kgg} into \eq{eq0}:
$$
	\rho(x)=c g'(x) g(x)^\alpha ( N \langle g\rangle)^{\alpha+1}.
$$
Let us remark that the assumptions on $k(x)$ forces
$g(x)$ to be non-decreasing with $g(\infty)>g(0)>0$.

In the case $\alpha\neq-1$, the normalization condition $R(\infty)=1$
results in
\begin{equation}
	\rho(x) =
	  \frac{\alpha+1}{g(\infty)^{\alpha+1}-g(0)^{\alpha+1}}
	   g'(x) g(x)^\alpha.
	\label{rhogg} 
\end{equation}
By multiplying both sides of the previous equation by $g(x)$ and
integrating from $0$ to $\infty$, one gets an expression for
$\langle g\rangle$ that does not explicitly contain the function $\rho(x)$.
This expression can be used to get the allowed value of the constant
$c$:
$$
	c = \frac 1 {N^{\alpha+1}}
		\left[\frac{g_\infty^{\alpha+1}-g_0^{\alpha+1}}{{\alpha+1}}\right]^{\alpha}
		\left[\frac{{\alpha+2}}{g_\infty^{\alpha+2}-g_0^{\alpha+2}}\right]^{\alpha+1}
$$
if $\alpha\neq-2$, with $g_\infty=g(\infty)$ and $g_0=g(0)$.
The particular cases $\alpha=\{-1,-2\}$ can be similarly treated.

The case $f(x,y)=f(x-y)$ is more complicated.
In this case both the nearest neighbor connectivity and clustering
coefficient depend on the fitness $x$ and conversely on the degree
$k$.
We managed to solve this case in the particular case of an
exponentially distributed fitness.
We indicate with $F(x)$ the rhs of \eq{Ieq}.
Thus \eq{Ieq} becomes:
$$
   \int_0^\infty \!\!\! f(x-u) \rho(u) dx = F(x)/N.
$$
By changing the integration variable into $z=x-u$ we get:
$$
	\int_{-\infty}^x \!\!\! \rho(x-z) f(z) dz = F(x)/N
$$
that in the special case $\rho(x)=e^{-x}$ becomes:
$$
	\int_{-\infty}^x \!\!\! e^z f(z) dz = e^x F(x)/N.
$$
By differentiating with respect to the variable $x$ we finally obtain:
\begin{equation}
	f(x,y) = \frac{F(x-y)+F'(x-y)}{N}.
	\label{F+F}
\end{equation}
\begin{figure}
  \includegraphics[width=7cm]{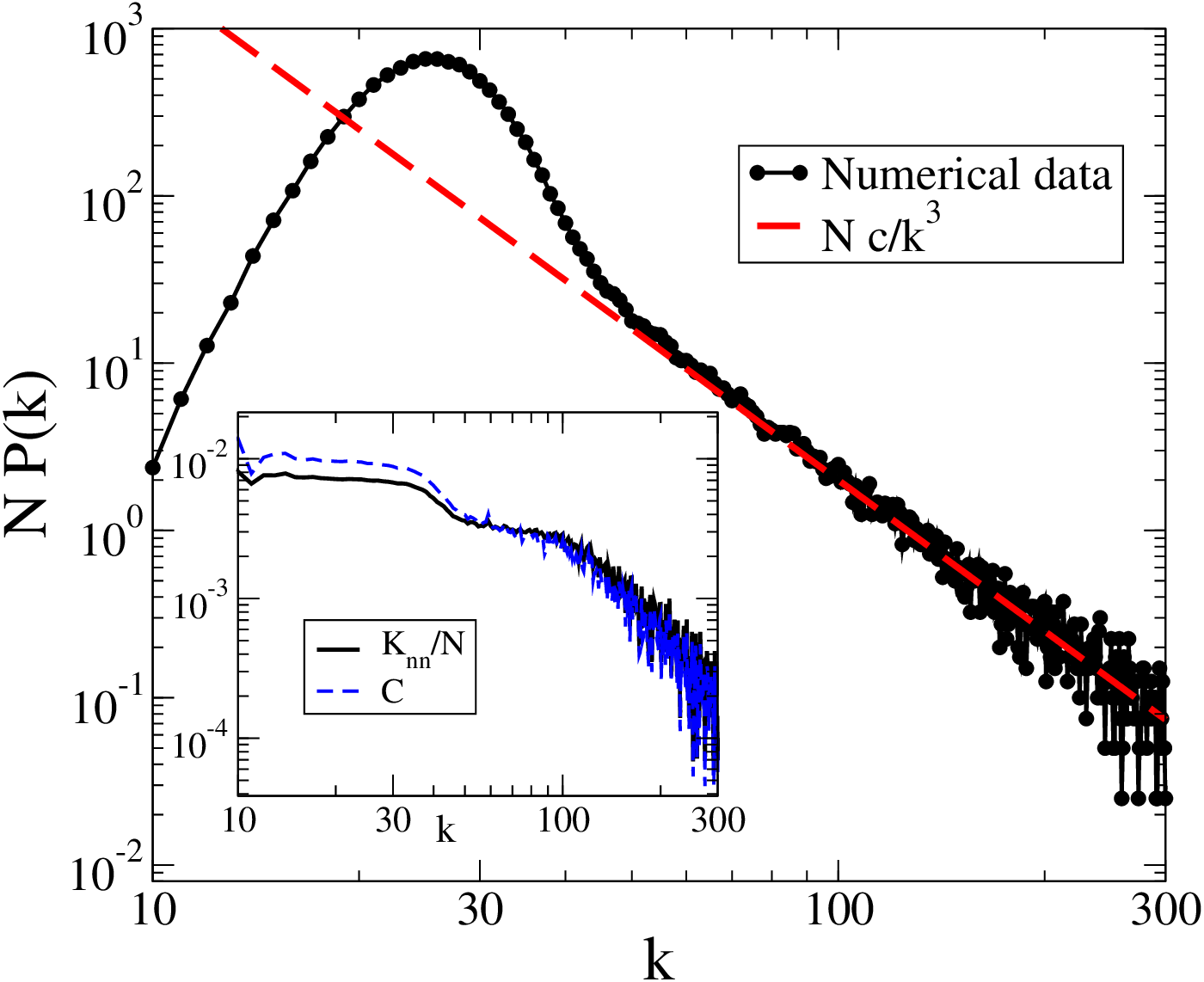}%
  \caption{Degree distribution in the case $f(x,y)=f(x-y)$,
	$\alpha=-3$, $\rho(x)=e^{-x}$, $k_0=10$, $N=10^4$,
	$f(u)=F(u)+F'(u)$ with $F(x)$ given by the lhs of \eq{Ieq}
	averaged $40$ times.
	   The value of $c$ is calculated from \eq{cnorm}.
	The inset shows the vertex degree correlation and transitivity
	as functions of the vertex degree. 
	\label{fig:fx-y}}
\end{figure}%
In order to test the result we take the function and parameter choice
of Fig.~\ref{fig:fx-y} caption.

The case $f(x,y)=f(x+y)$ is analogous.
%
\begin{figure}
  \includegraphics[width=7cm]{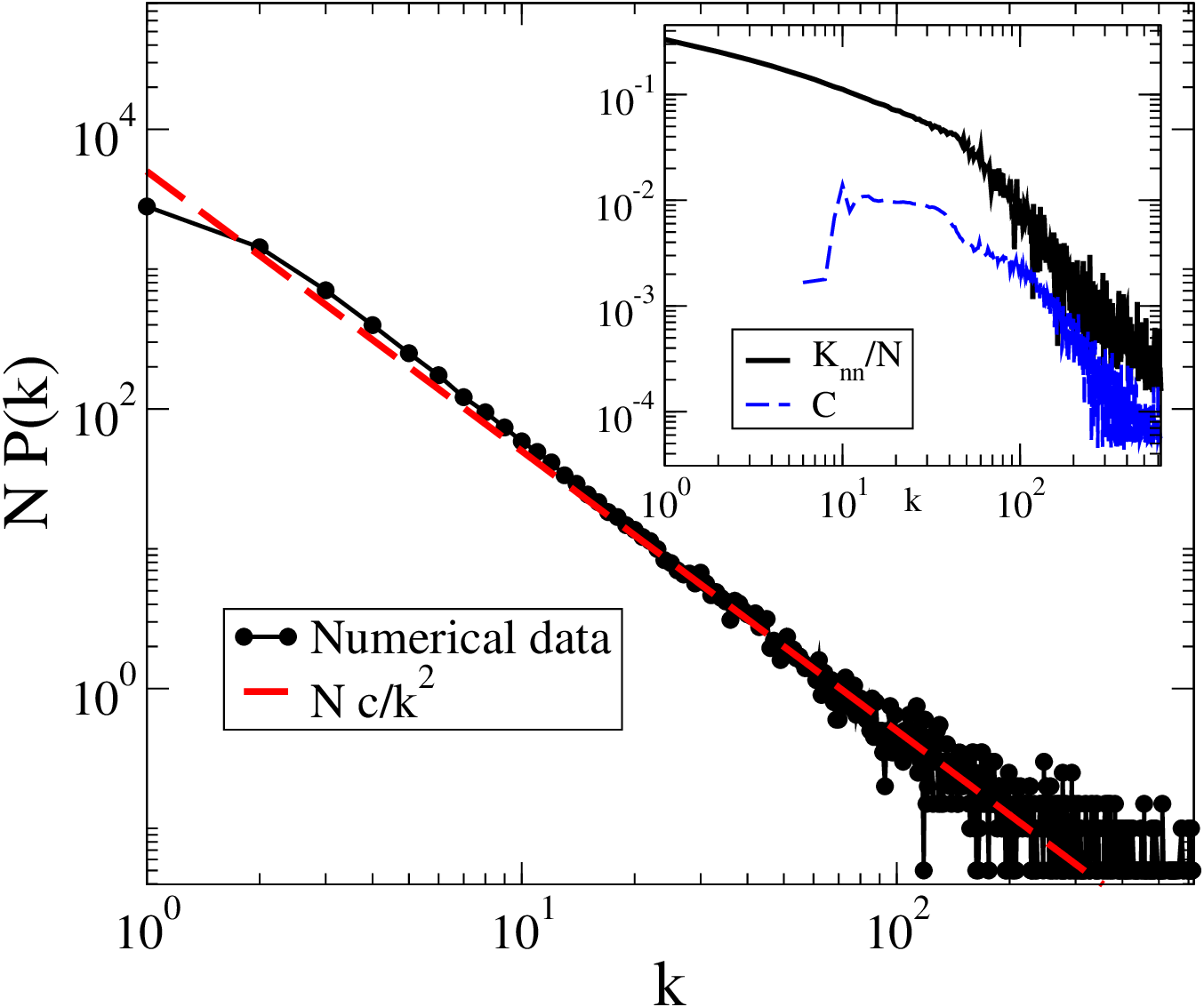}%
  \caption{Degree distribution in the case $f(x,y)=f(x+y)$ and 
	$\alpha=-2$, $\rho(x)=e^{-x}$, $k_0=0.5$, $N=10^4$,
	$f(v)$ from \eq{cont},
	averaged $20$ times.
	   The value of $c$ is calculated from \eq{cnorm}.
	The inset shows the vertex degree correlation and transitivity
	as functions of the vertex degree. 
		\label{fig:fx+y}}
\end{figure}%
Again, we consider the special case $\rho(x)=e^{-x}$, getting now:
\begin{equation}
	f(x,y) = \frac{F(x+y)-F'(x+y)}{N}.
	\label{F+F2}
\end{equation}
The solution of \eq{Ieq} for $\alpha=-2$ obtained via \eq{F+F2} reads,
recalling that \(k_0=\beta N\) and using \eq{cnorm},
\begin{equation}
	 f(x,y) = 
		\frac {1}
		{[1-(\beta^{-1}-1)e^{-(x+y)}]^2}.
	\label{cont}
\end{equation}
Through \eq{cont} we clarify the assumption made in the
original paper by Caldarelli \etall \cite{gcalda}, where 
\mbox{$f(x,y)=\Theta(x+y-z)$} with $z=z(N)$.
Note that now with the latter choice of $f(x,y)$ one gets $P(k)=N e^{-z}
k^{-2}$ that forces $z$ to depend upon $N$ in order to get the
correct normalization.
The functional form of the $z(N)$ was numerically guessed by
Ref.~\cite{millozzi}.
 To test the result we take the parameters reported in the
caption of Fig.~\ref{fig:fx+y}.

In these last two cases, both the nearest neighbor connectivity and
clustering coefficient show non trivial $k$ dependence.

In conclusion we present a general procedure to reproduce real SF
networks with arbitrary vertex degree distribution densities.
More specifically,
we found that, given a fitness distribution density $\rho(x)$,
it is always possible to find a symmetric linking probability function
$f(x,y)$ such that the resulting random network is scale-free with a
given real exponent.
We give the recipe to find these linking functions, in three cases of
interest.
In order to allow the generation of networks even closer to the real
data, 
it would be desirable to have control not only on the vertex degree
distribution, but also on the vertex transitivity
and vertex degree correlation, by solving simultaneously
Eqs.~(\ref{Knn}), (\ref{C}), (\ref{Pkx}).
As a first step, the compatibility of these three equations should be
addressed, once the functions $P(k)$, $K_{\mathrm{nn}}(k)$, $C(k)$ are
given.
The solution of this problem is certainly very hard and is left open
for the future.
The relative ease with which we obtain SF structures seems to be the
key ingredient in order to explain the ubiquitous presence and
robustness of the real data.

We acknowledge R.~Pastor-Satorras, P.~De~Los~Rios, D.~Garlaschelli,
F.~Squartini,
S.~Millozzi, S.~Leonardi for useful discussions and support
from EU FET Open Project IST-2001-33555 COSIN (www.cosin.org).
VDPS and GC thank IMFM PAIS PA\_G02\_4 for support.

\end{document}